\documentclass[twocolumn,showpacs,preprintnumbers,amsmath,amssymb,prb]{revtex4}
\usepackage{wasysym}
\usepackage{graphicx}
\usepackage{dcolumn}
\usepackage{bm}
\usepackage{amssymb}
\usepackage{epsfig}
\usepackage{graphicx,amsmath,amssymb,bm}
\newcommand{\be}{\begin{equation}}
\newcommand{\ee}{\end{equation}}
\newcommand\beq{\begin{eqnarray}}
\newcommand\eeq{\end{eqnarray}}

\begin{document}
\title{Finite size effects on the statistical angle of an electron induced dyon in proximity to a topological insulator}

\author{Sichun Sun}
\affiliation{Institute for Nuclear Theory, Box 351550, Seattle, WA 98195-1550, USA}
\author{Andreas Karch}
\affiliation{Department of Physics, University of Washington, Seattle, WA 98195-1560, USA}

\date{\today}
\preprint{INT-PUB-11-033}

\begin{abstract} A pointlike electric
charge close to the surface of a three dimensional topological insulator induces a magnetic monopole mirror charge. We study the distance dependence of the statistical angle describing
this induced dyon system.
We find that the total angular momentum, which sometimes is used as signature of the statistical angle, for an electron outside a
finite size spherical or tube shaped
topological insulator is zero
for any finite distance between the electron and the surface. However, we show that in the 2-electron system one can indeed isolate a non-trivial statistical angle for intermediate size loops, that is loops much larger than the distance of the charge to the sample but much smaller than the size of the sample. The necessity for this limit confirms the 2+1 dimensional nature of the non-trivial
statistical angle found in previous work.
Our results clarify the conditions under which the statistical angle of this system could be measured in real experiments. \end{abstract}

\date{\today}
\maketitle

\section{Introduction} The low energy effective theory describing three
dimensional topological insulators (TIs) is given by Maxwell
electromagnetism augmented by an axion like $\vec{E} \cdot \vec{B}$ term
\cite{qi2008} leading to modified constitutive relations. For the bulk
material this effective theory is valid at energy scales below the gap. In
the presence of interfaces the massless modes on the surface would have to
be included in the effective description of the material and the low
energy effective description in terms of Maxwell theory with modified
constitutive relations only applies if the surface modes are gapped by an
external (time reversal breaking) perturbation. Furthermore the
effective description is only valid at energy scales below this
induced surface gap. Such
an external breaking can easily be set up experimentally, e.g. by a magnetic field.
Several potential experimental consequences follow from this effective
theory, such as, for an example, a non-trivial Faraday and Kerr rotation
\cite{qi2008}. By scanning the external field, the topological
contribution to the Faraday effect from the time reversal breaking field can be cleanly
separated from the topological contribution. One of the most
spectacular predictions of this effective theory is the appearance of a
magnetic monopole mirror charge when solving for the static
electromagnetic fields sourced by a single point charge (located
inside a topological trivial insulator such as e.g. vacuum) in the
presence of
a TI interface \cite{2009Sci...323.1184Q}.

For an infinitely extended planar interface, the corresponding magnetic
mirror charge is a pointlike magnetic monopole (also carrying some electric charge). As always with mirror
charges, this monopole of course is not physical but simply a
mathematical tool
to calculate the magnetic fields in the physical region (that is
for calculating the fields above the interface, the monopole appears
to be located
below the interface and vice versa).
Microscopically it is surface currents on the interface that source a
magnetic field with a $1/r^2$ fall-off where $r$ is the distance to the
mirror monopole. Nevertheless, the magnetic fields generated this way are,
in the physical region, indistinguishable from the ones generated by a
genuine monopole and so share some of its properties. In particular, it is
well known that the electromagnetic fields generated by an electric point
charge $e$ and a spatially separated monopole of magnetic charge $g$ carry
a net angular momentum which has several interesting properties: it
is independent of the distance between
charge and monopole, pointing in the direction from the charge to the
monopole and proportional to $e g$ (see e.g. Ref.~\onlinecite{jackson}). This total angular momentum of
the composite dyon formed by the charge-monopole pair is the sum of the angular momenta of the two point particles and this
angular momentum stored in their fields. Via the spin-statistics theorem this shift in the angular
momentum of the dyon is often interpreted as a shift in the statistical angle that determines the
behavior of the multi-dyon wavefunction under the exchange of two dyons. For a genuine charge/monopole pair, the Dirac quantization of magnetic
charges ensures that the resulting angular momentum is an integer multiple
of $\hbar/2$. So while the statistics of the dyonic system can be changed
due to angular momentum stored in the fields, the net angular
momentum is still properly quantized and so the overall statistical angle is always
an integer multiple of $\pi$. The full dyonic
system is either a fermion or a
boson.

As the field of the mirror monopole is indistinguishable from the field of a real monopole, the same
calculation implies that the electromagnetic fields generated by a
single electric point charge $e$ above the interface of a TI carry a
non-trivial angular momentum as well. As the
electric as well as the magnetic field
below the interface (that is inside the TI)
appear to be sourced by coincident charges located at the location of the
actual physical charge, the contribution to the angular momentum from
that region of space vanishes. By symmetry, the actual angular momentum
in the system obtained from integrating over the electric and magnetic
fields above the interface, sourced by the physical electric charge as
well as the mirror charge, gives exactly half of the angular momentum
one would get from a genuine charge/monopole pair with the same values
of $e$ and $g$ (for details see the appendix). However the magnitude of the
induced mirror charge is proportional to the finestructure constant
$\alpha$ and furthermore depends continuously on the material properties $\mu$ and
$\epsilon$. Consequently, it generically does not obey Dirac quantization
conditions. The resulting angular momentum is not
quantized. As a result, the statistics one associates with these charge carriers based on their angular momentum is no
longer simply fermionic or bosonic. Instead they seem to behave as anyons \cite{2009Sci...323.1184Q} with a
statistical angle given by \footnote{Our expression differs by a factor
of $2 \pi$ from the one quoted in \cite{2009Sci...323.1184Q}. A factor
of $4 \pi$ seems to be due to the fact that in \cite{2009Sci...323.1184Q},
which works in Gaussian units, the formula for the statistical angle
has been normalized to yield $\theta_S = \pi$ for a monopole
of flux $hc/e$. However latter has been taken to correspond to
$g=hc/e$, which would be correct if $\vec{\nabla} \cdot \vec{B}
= g \delta(\vec{r})$ as it is in SI units. However in Gaussian
units $\vec{\nabla} \cdot \vec{B}
= 4 \pi g \delta(\vec{r})$ and so the formula for $\theta_S$ in
\cite{2009Sci...323.1184Q} in terms of general $e$ and $g$ should be modified by a factor of $4 \pi$. The remaining factor of 2 mismatch seems
to be due to the fact that in \cite{2009Sci...323.1184Q} the factor
of 1/2 from the fact that only the half of space above
the interface contributes to the angular momentum has not been taken
into account. For the reader's convenience we rederive
the expression of the statistical angle in SI units in the appendix,
where intermediate steps can be readily compared to standard textbook
expressions from e.g. \cite{jackson}. This confirms the factor of $2 \pi$
as we have it here. For the bulk of the paper we follow
\cite{2009Sci...323.1184Q} and use Gaussian units. The final expression
for the statistical angle has to be independent of choices of units
when only expressed in terms of the fine structure constant $\alpha$ as well
as the ratios $\epsilon/\epsilon_0$ and $\mu/\mu_0$.}
\beq
\theta_S=\pi \frac{L}{\hbar}= \frac{4 \pi \alpha^2
P_3}{(\frac{\epsilon_1}{\epsilon_0}+\frac{\epsilon_2}{\epsilon_0})
(\frac{\mu_0}{\mu_1}+\frac{\mu_0}{\mu_2})+4\alpha^2P_3^2} \label{thetas}.\eeq
$P_3=\theta/(2 \pi)$ is the electromagnetic polarization. It is
$0$ in a topologically trivial material and $1/2$ inside a TI. In
order to ensure that our expression for $\theta_S$ can be compared
between different unit systems, we explicitly displayed
factors of $\epsilon_0$ and $\mu_0$. For
the bulk of this work we'll work with units where $\epsilon_0=\mu_0=1$.

The non-trivial statistical angle was interpreted in Ref.~\onlinecite{2009Sci...323.1184Q}
as a result of the two dimensional nature of the TI surface. It is well
known that in two spatial dimensions, anyonic statistic is allowed.
The authors of Ref.~\onlinecite{2009Sci...323.1184Q} proposed that in the vicinity
of a TI surface an electric point charge indeed turns into an anyon
with statistical angle $\theta_S$ and proposed an explicit experimental
setup that would allow its measurement. This proposal raises
one important conceptual puzzle: the angular momentum in the system and hence the inferred
$\theta_S$ is entirely independent
of the distance $z_0$ between the point charge and the interface.
While it is reasonable to assume that a point charge in close vicinity
of a TI surface has anyonic character, the calculation of $\theta_S$ via the induced angular momentum
seems to predict that any point charge moving freely in three dimensional
space would pick up a statistical angle $\theta_S$ provided there is
a planar TI interface somewhere in the universe at arbitrary large
distance $z_0$, which is very counter-intuitive and also
seems to indicate that the anyon is truly 3+1 dimensional in character,
contradicting the fact that 3+1 dimensional anyons should be
impossible (the exception recently proposed in Ref.~\onlinecite{teokane} can
readily be understood in terms of a more complicated topology
of configuration space in this case \cite{nayak}).

One could expect the $z_0$-independence of the angular momentum
to be an artifact of the special example of an infinite planar interface.
After all, $L/\hbar$ is dimensionless and so could only depend on $z_0$
in the form of the ratio $z_0/a$, where $a$ is another geometric
scale in the problem. For the infinite plane, no such other scale is present.
With this puzzle in mind, we analyze the angular momentum associated
with an electric point charge in the vicinity of a TI interface for two
different geometries: a TI in the shape of a sphere and a TI
with a semi-infinite tube like shape inside a perfectly conducting
cavity. We find that in {\it both} cases the angular momentum
vanishes identically for any charge separated from the surface even by an
infinitesimal amount. These two examples make us suspect that the angular momentum
will in fact vanish for a charge close to (but not right on top) the
surface of any finite size TI. In hind-sight, this result is not too surprising. After all,
the microscopic description of the topological insulator is in terms of a system of
electrons and protons with properly quantized charges, obeying the standard rules of quantum
mechanics. Any state described by this microscopic system has to have a properly quantized
angular momentum. As long as the effective theory correctly captures the long distance behavior of the system,
it has to obey the quantization conditions obeyed by the microscopic constituents. So the angular momentum $L$ for a charge close to any compact TI has to be an integer multiple of $\hbar/2$. As a consequence, $L$ can not continuously vary as a function of $z_0/a$. As we expect $L \rightarrow 0$ for $z_0/a \rightarrow 0$, it should have been expected that $L=0$ is indeed the correct answer for all finite values of $z_0/a$.

At first this result may indicate that the non-trivial $\theta_S$ identified in the planar case does not carry over to any compact sample and hence would not be measurable. But this is too naive. After all, total angular momentum of the one-particle system was only taken as a stand-in for the statistical angle of the excitations. A more careful analysis should directly analyze the two-particle system and study the change in action associated with a non-trivial loop in configuration space. Performing this analysis we find in the case of the planar interface that $\theta_S$ as inferred from the angular momentum in the one-particle system only describes the exchange of two particles in the limit that the size of the loop $l$ is much larger than the separations $z_0^{(1)}$ and $z_0^{(2)}$ of the charges to the surface of the TI. This is  consistent with the interpretation of $\theta_S$ as a topological effect. For $l \sim z_0^{(1),(2)}$ short distance effects become important. In this limit details of the path matter. But for $l \gg d_{1,2}$ the only effect surviving is the topological phase (which in this limit is independent of the shape of the loop as it
should be). So the statistical angle governs large loops in configuration space.
However, for the realistic case of a compact TI of linear size $a$, we should clearly expect significant finite size effects in the case that loops are of order the sample size, $l \sim a$. Indeed this expectation is born out. For generic $z_0^{(1),(2)}$ and $a$ the change in action associated to taking particle 2 around particle 1 depends crucially on the path and has no relation to the $\theta_S$ obtained from the planar case. However, in the intermediate loop size regime
$$ z_0^{(1),(2)} \ll l \ll a$$
we once more are able to show that the phase is topological (independent of shape) and is given by the flat space value $\theta_S$. While not surprising, this analysis clearly lays out that any experimental attempt at measuring $\theta_S$ e.g. as proposed in Ref.~\onlinecite{2009Sci...323.1184Q} has to be set up as to operate in this intermediate loop size regime.

This note is organized as follows: In the next section we'll
review the solution for the potentials in the case of a spherical
TI and construct the solution of the potentials in the case of
a tube-like TI. In section 3 we analyze the distance dependence of the statistical angle. We first show that in the one particle system angular momentum vanishes identically in both examples. We then analyze explicitly the two-particle
 system and show, in the case of the sphere, that $\theta_S$ obtained from the flat space analysis does govern the intermediate loop size regime. We discuss the relevance
of these results in section 4.

\section{Electric and Magnetic Potentials in two compact examples}
\subsection{Spherical TI}

\begin{figure}[t] \includegraphics[width=8
cm]{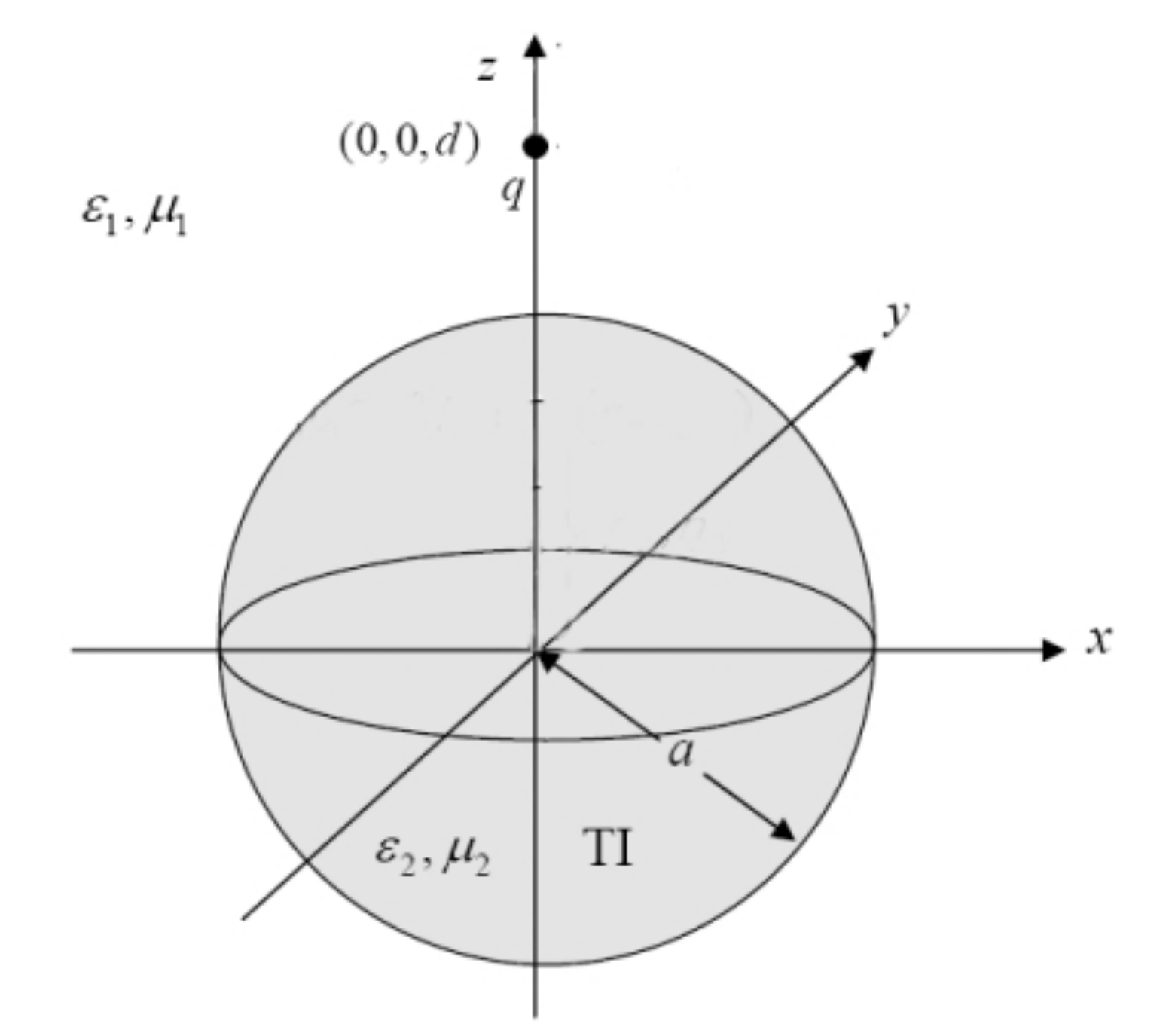}
\caption{\label{fig:sphere} A charge $q$ outside a spherical TI. }
\end{figure}

First let us analyze the electric and magnetic fields for
a spherical
topological insulator and a point-like electric charge outside
the sphere. The corresponding potentials have been worked out
in the supplementary material of Ref.~\onlinecite{2009Sci...323.1184Q}.
As shown in Fig.~\ref{fig:sphere}, a spherical topological
insulator of a radius $a$ and a magneto-electric polarization $P_3$ is
centered at the origin, and a point-like electric charge is at $(0,0,d)$.
$\epsilon_1$ and $\mu_1$ are
the dielectric constant and the magnetic permeability
outside the sphere, $\epsilon_2$ and $\mu_2$ the corresponding
quantities inside the sphere.\\ Both
inside and outside the sphere, the curl of electric and magnetic fields is
zero, thus we can find a scalar potentials in both regions:
\beq \nonumber&E^{(i)}=-\triangledown \Phi_E^{(i)}\\
&B^{(i)}=-\triangledown \Phi_M^{(i)}
\label{potentials}
\eeq
where $i=1,2$ stand for inside
and outside region. The most general solution for the potentials
in eq.~\eqref{potentials} can be written in terms of Legendre
polynomials:
\beq \nonumber&\Phi_E^{(1)}= \frac{q}{\epsilon_1}\sum
\frac{r^l}{d^{l+1}}P_l(\cos\theta) + \sum A_l
(\frac{a}{r})^{l+1}P_l(\cos\theta)\\ \nonumber&\Phi_E^{(2)}= \sum B_l
(\frac{r}{a})^{l}P_l(\cos\theta)\\ \nonumber&\Phi_M^{(1)}= \sum C_l
(\frac{a}{r})^{l+1}P_l(\cos\theta)\\ &\Phi_M^{(2)}= \sum D_l
(\frac{r}{a})^{l}P_l(\cos\theta) \label{spherepotentials} \eeq
Solving boundary condition for
the interface between trivial and topological insulator
(that is continuity of the perpendicular components
of $\vec{D}$ and $\vec{B}$ as well as the parallel components
of $\vec{H}$ and $\vec{E}$), one arrives at:
\beq &\nonumber A_l=
\frac{q}{\epsilon_1}\frac{a^l}{d^{l+1}}[\frac{(\epsilon_1l-\epsilon_2l)[l/\mu_1+(l+1)/\mu_2]-(2\alpha
P_3)^2l(l+1)}{(2\alpha
P_3)^2l(l+1)+(\epsilon_1(l+1)+\epsilon_2l)[l/\mu_1+(l+1)/\mu_2]}]\\
&\nonumber B_l=
\frac{q}{\epsilon_1}\frac{a^l}{d^{l+1}}[\frac{(\epsilon_1l-\epsilon_2l)[l/\mu_1+(l+1)/\mu_2]-(2\alpha
P_3)^2l(l+1)}{(2\alpha
P_3)^2l(l+1)+(\epsilon_1(l+1)+\epsilon_2l)[l/\mu_1+(l+1)/\mu_2]}+1]\\
&\nonumber C_l= q\frac{a^l}{d^{l+1}}\frac{(2\alpha P_3)^2l(2l+1)}{(2\alpha
P_3)^2l(l+1)+(\epsilon_1(l+1)+\epsilon_2l)[l/\mu_1+(l+1)/\mu_2]}\\
& D_l= q\frac{a^l}{d^{l+1}}\frac{-(2\alpha
P_3)^2(l+1)(2l+1)}{(2\alpha
P_3)^2l(l+1)+(\epsilon_1(l+1)+\epsilon_2l)[l/\mu_1+(l+1)/\mu_2]}
\label{spherecoefficients} \eeq
The fields here could be considered to be generated by a point image
electric charge, magnetic monopole, and a line of image electric or
magnetic charges\cite{2009Sci...323.1184Q}.

\subsection{Semi-infinite rectangular TI-tube in conducting cavity}

\begin{figure}[t]
\includegraphics[width=8 cm]{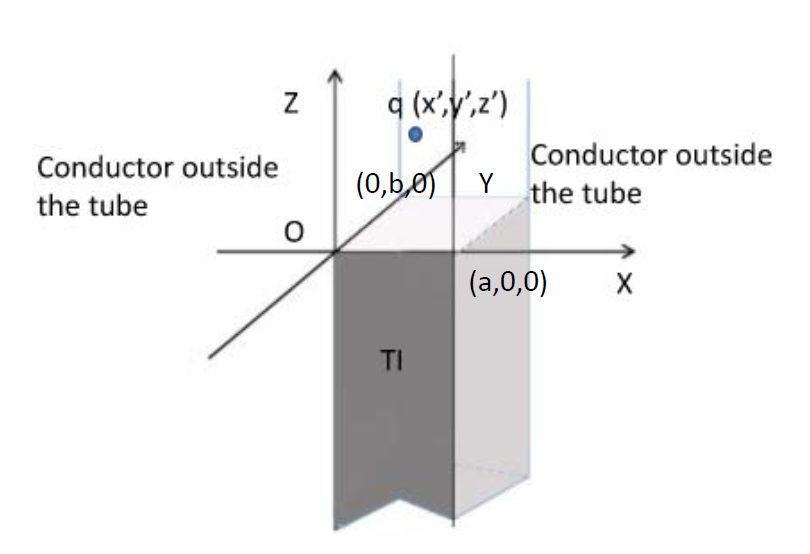}
\caption{\label{fig:tube} An electric charge $q$ inside the
 positive half-tube $z>0$ with dielectric constant $\epsilon_1$ and
magnetic permeability $\mu_1$ in the vicinity of a TI
with $\epsilon_2$  and $\mu_2$ filling the negative half-tube $z<0$.
The tube has a width $a$, and length $b$. We assume that the walls of the
tube are formed by a perfect
conductor.}
\end{figure}

Next we want to consider the case of a
semi-infinite topological insulator (the TI being
at $z<0$) tube with rectangular
cross section inside a conducting
wall, see Fig.~\ref{fig:tube}. An
electric point-like charge is located at $(x',y',z')$. As in
the spherical
case, we can write down general forms for electric and magnetic
potential:
\beq \nonumber&\Phi_E^{(1)}= \frac{q}{\epsilon_1}\sum O_{mn}
\sin(k_m x) \sin(k_n y)e^{\gamma_{mn}(z-z')} \\
\nonumber&+ \sum A_{mn} \sin(k_m x) \sin(k_n
y)e^{-\gamma_{mn}z}\\ \nonumber&\Phi_E^{(2)}= \sum B_{mn} \sin(k_m
x) \sin(k_n y)e^{\gamma_{mn}z}\\ \nonumber&\Phi_M^{(1)}= \sum C_{mn} \sin(k_m
x) \sin(k_n y)e^{-\gamma_{mn}z}\\ &\Phi_M^{(2)}= \sum D_{mn} \sin(k_m
x) \sin(k_n y)e^{\gamma_{mn}z}
\label{tubepotentials}
\eeq
with $O_{mn}$ being the coefficients of the Green's
function of a point-like charge in this system:
\beq
O_{mn}=\frac{2}{ab\gamma_{mn}}\sin(k_m x') \sin(k_n y'). \eeq
Solving boundary
condition, we arrive at:
\begin{eqnarray}
\nonumber D_{mn} &=&
-\frac{2qO_{mn}e^{-\gamma_{mn}z'}(2\alpha
P_3)}{(\epsilon_1+\epsilon_2)(1/\mu_1+1/\mu_2)+(2\alpha P_3)^2}\\
\nonumber C_{mn} &=& -D_{mn}\\
 B_{mn} &=&
\frac{2qO_{mn}e^{-\gamma_{mn}z'}(1/\mu_1+1/\mu_2)}{(\epsilon_1+\epsilon_2)(1/\mu_1+1/\mu_2)+(2\alpha
P_3)^2}
\label{tubecoefficients}
\\
 \nonumber A_{mn} &=& q
O_{mn}e^{-\gamma_{mn}z'}
\\
 & &(-\frac{1}{\epsilon_1}+\frac{2(1/\mu_1+1/\mu_2)}{(\epsilon_1+\epsilon_2)(1/\mu_1+1/\mu_2)+(2\alpha
P_3)^2})
\nonumber
\end{eqnarray}

In this case the geometry of the TI is not really compact. However, due to the Dirichlet boundary conditions on the conducting walls from the 1d point of view all our gauge fields are massive and exponentially decay at large $|z|$. For all practical purposes, the TI is finite in extend in the $z$-direction.

\section{Distance dependence of statistical angle}
\subsection{Total angular momentum as a global probe of the system}

The
total angular momentum is given by: \beq L=\int
\vec{x}\times \frac{(\vec{E}\times \vec{H})}{4 \pi c^2} d^3x=\frac{\epsilon}{4\pi}\int
\vec{x}\times(\vec{E}\times \vec{B})d^3x. \label{l} \eeq
The extra factor of $1/(4 \pi)$
compared to Ref.~\onlinecite{jackson} is due to the fact that we
follow Gaussian units as in Ref.~\onlinecite{2009Sci...323.1184Q}.
In a normal system without a
topological insulator, there would not be any
non-zero angular-momentum of the
system in the presence of only a static point-like electric charge.
In the presence of a topological insulator, $\vec{L}$ can be non-zero.
In order to address the question about distance dependence of the angular momentum and the statistical angle in the case of TIs with a compact geometry, we want to calculate the angular momentum in the two examples described in the previous section.

Let us start with the case of the
rectangular tube. Plugging our answers
from eqs.~\eqref{potentials}, \eqref{tubepotentials} and \eqref{tubecoefficients}
into the expression for the angular momentum eq.~\eqref{l},
we see that the integral over $x$ and $y$
can easily be done
analytically. For example, starting with the $x$ momentum density
\beq
\nonumber &l_x = y (\partial_x \Phi_E \partial_y \phi_M - \partial_y \Phi_E \partial_x \phi_M)\\
& - z (\partial_z \Phi_e \partial_x \phi_M - \partial_x \phi_E \partial_y \phi_M)
\eeq
we see that integrating over $y$ first gives an expression proportional to
\beq \nonumber&\int_0^b \,  dy \, l_x \sim m_1 \cos(m_1 \pi x/a) \sin(m \pi x/a)\\
&+ m \cos(m \pi x/a) \sin(m_1 \pi x/a)\eeq
for integers $m$, $m_1$ still to be summed over (as well as $n$ and $n_1$ appearing in the coefficients). Further integration of this expression over $dx$ then vanishes identically. We have confirmed that $L_y$ and $L_z$ similarly vanish after doing the $x$ and $y$ integrals using Mathematica.

The calculation in the spherical case is a little more cumbersome. Plugging eqs.~\eqref{potentials}, \eqref{spherepotentials} and
\eqref{spherecoefficients} into eq.~\eqref{l}, we can use the recurrence
relation of Legendre polynomials
to simplify the integrand.
After performing the $\theta$ integral, the double sums occurring in $\vec{E}
\times
\vec{B}$ collapse into a single one.
Performing the integral in spherical
coordinates, we
once more find that the angular momentum vanishes identically as
long as the electric charge is separated even an infinitesimal amount
from the surface of the sphere, irrespective of the distance $d$.
If we set the distance $d$ to zero to begin with and then perform the integrals and sums,
we get back to the result of eq.~\eqref{thetas} valid for the infinite half-plane.
The fact that for $d=0$ the sums give back this non-trivial
result is a non-trivial check. For $d/a=0$ one should clearly recover
the result of the plane, which can also be thought of as $a \rightarrow
\infty$. What is surprising is that this limit is not smooth. The angular
momentum vanishes for any finite $d$.\\
As discussed in the introduction, this result should have been expected based on microscopic considerations. For a real finite system made of electrons and protons total angular momentum has to be quantized. So it can not continuously depend on $d/a$. For it to vanish at infinity it has to be zero for all $d/a$. Even for the topological insulator, genuine fractional angular momentum should only exist in the infinite system. With the electron charge being on the surface, one can get back the infinite system result.

\subsection{Finite size loop path in the two particle system}

While the angular momentum for a genuine charge/monopole pair correctly captures the statistical angle, our zero result from the previous subsection strongly suggests that in the case of a spatially compact TI one should be more careful. To directly obtain the statistical angle, the natural thing to do is to look at two particles moving adiabatically around each other and study the phase change of the action. The action of two point particles at a fixed distance $z_0$ above a TI/insulator interface located at $z=0$ contains the standard terms coupling the point particles to the gauge fields:
\beq
\label{ptaction}
S = e \int A^{(2)}_{\mu} \frac{d x_{(1)}^{\mu}}{d \tau} + (1 \leftrightarrow 2)
\eeq
where the super/sub-scripts $(1)$ and $(2)$ refer to the two particles, $A_{\mu}^{(i)}$ being the field sourced  by particle $i$, $x^{\mu}_{(i)}$ its position. Let us for simplicity look at the case where particle 1 is kept at the origin in the $x$-$y$ plane and particle 2 is taken around a non-trivial loop.

\subsubsection{Planar interface}
It is easy to work out the effect of this coupling for the planar interface in detail. Note that we can write the magnetic field of the mirror monopole of magnetic charge $g$ at $z=-z_0$ induced by an electric charge $e$ at $z=z_0$ (and bothat at $x=y=0$) in terms of a vector potential:
$$ A^{(1)}_{\phi} = \frac{g}{4 \pi} ( 1- \cos \theta)$$
where $\phi$ and $\theta$ are the angles in a spherical coordinate system centered on the location of
the mirror monopole at $(0,0,-z_0)$. That is the 2nd particle at $(x_{(2)}, y_{(2)},z_0)$ in this coordinate system is located at $\tan \theta_{(2)} = \rho_{(2)}/z_0$ where $\rho^2=x^2+y^2$. This is the standard form of the vector
potential of a monopole with a Dirac string running along the negative $z$-axis. This is the appropriate form to use for a mirror charge located below the interface, as this mirror charge is only supposed to be used when calculating fields above the interface (for the magnetic field below the interface one would similarly use the $\vec{A}$ associated to a monopole with a Dirac string running along the positive $z$-axis for a monopole located at $z=+z_0$. As there is no physical charge located below the interface and so no contribution to eq.~\eqref{ptaction} we will not need $\vec{A}$ in this region).

If we keep one particle fixed at $(x=0,y=0,z=z_0)$ and take the other particle around a loop at a fixed $z_0$  ($\phi_{(2)}$ goes to $\phi_{(2)} + 2 \pi$ at fixed $\theta_{(2)}$), $e^{iS}$ picks up a phase:
$$
e^{i S} \rightarrow e^{iS} e^{e\, \int d \tau \dot{\phi}_{(2)} A^{(1)}_{\phi}} = e^{i \frac{e\, g}{2} (1- \cos \theta_{(2)}) }
$$
We see that due to the $\theta_{(2)}$ dependence already in the case of the planar interface, the resulting statistical angle depends on path. As $\tan \theta_{(2)}=\rho_{(2)}/z_0$, we get different answers depending on the size of the loop. When the charge is a finite distance $z_0$ above the interface and taken around a very small loop of size $l \ll z_0$ we have $\theta_{(2)}=0$ along the whole path
and correspondingly do not pick up any phase. With a loop large compared to $z_0$ (that is $l \gg z_0$), $\theta_{(2)} \rightarrow \pi/2$. In this limit we get a phase  $e^{i e\, g/2}$ in the action independent of the detailed shape of the path in the $x$-$y$ plane \footnote{These answers also obvious when one considers that $\vec{\nabla} \times \vec{A}= \vec{B}$, so the line integral of $\vec{A}$ we are doing is equal to the total magnetic flux through the surface enclosed by the loop. For $\theta_{(2)}=0$ the loop has vanishing area and no flux. For $\theta_{(2)}=\pi/2$ the loop captures all the flux in the northern hemisphere.}. This change in the action corresponds to $e^{2 i \theta_S}$ where $\theta_S$ is the statistical angle describing the exchange of two particles. Exchanging particle 1 and 2 twice should correspond to a closed loop in configuration space as the one we have been analyzing. So $\theta_S = e g/4$. Plugging in the $g$ obtained for the mirror charge in Ref.~\onlinecite{2009Sci...323.1184Q} (rederived in SI units in the appendix here) we do get back to eq.~\eqref{thetas}.

In conclusion, we find that for a planar interface in the realistic case of a finite $z_0$ (so that our effective theory applies), only a large loop (compared to $z_0$) gives a shape independent answer governed by a topological statistical angle. For these large loops the statistical angle is half of that of a dyon made of a real electron/monopole pair in complete agreement with the angular momentum calculation in previous section. The modification for small loops presumably should be understood as a result of short range interactions.\\

\subsubsection{Spherical TI:}
\begin{figure}[t]
\includegraphics[width=7 cm]{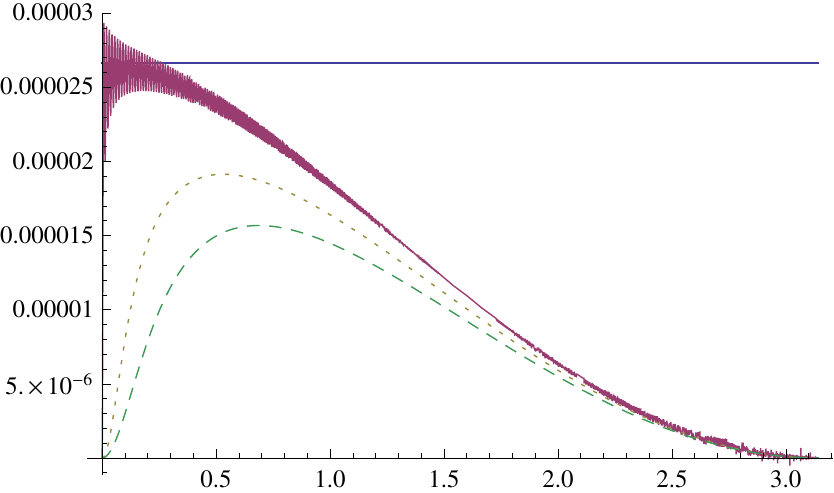}
\caption{\label{fig:plot} $A_\phi$ versus $\theta$ with different $z_0$'s for the mirror charge on an electron. $z_0=0.05a$ for the dotted line and $z_0=0.1a$ for the dashed line and $z_0=0.0005a$ for the solid curved line. The horizontal straight line stands for the universal angle $\theta_S$. The thickness of the $z_0=0.0005a$ line represents the numerical uncertainty due to the slow convergence of
the large $l$ truncation. In this plot we chose $a=1$ and all permittivities and permeabilities equal 1. The small numerical value of $\theta_S$ is due to the appearance of $\alpha$ in the expression for the mirror charge.}
\end{figure}

For the case of a spherical TI of radius $a$ we can also obtain the vector potential associated with a single point charge at $(0,0,d)$ with $d=a+z_0$. It again only has a $\phi$ component with
$$ A_{\phi} = - C_0 a \cos \theta - \sum_l^{\infty} \frac{C_l}{l} \frac{a^{l+1}}{r^l} \sin(\theta) P_l^1 (\cos \theta).$$
The $C_l$ are given in eq.~\eqref{spherecoefficients}. Most importantly, $C_0=0$ (this would be a net magnetic charge, so it has to vanish).\\
We use this expression for $A_\phi$ and study the phase change in spherical coordinates when two particles located at the same $r=d=a+z_0$ but different $\theta$ are exchanged. We take one particle fixed at $(0,0,d)$ while the other one goes from $\phi$ to $\phi+2\pi$ at a given $\theta$. The action picks up a phase $2\pi A_\phi$. We plot this value for different $z_0$ together with the universal $\theta_S$ in Fig.~\ref{fig:plot}.\\
There are a couple of interesting features about this plot. First, $A_\phi$ always vanishes for $\theta \rightarrow 0$ and $\theta \rightarrow \pi$, as expected after considering the planar case. \\
Second, for intermediate $\theta$ between 0 and $\pi/2$, there is a certain non-zero plateau region. For small $z_0=d-a$ (small compared to $a$), this plateau region repeats the $\theta$ behavior of planar case before finite size effect kicks in. In this limit the plateau value of $A_{\phi}$ goes back to planar case result, the universal $\theta_S$. \\
This can also be seen analytically from the large $a$ asymptotic behavior of $A_\phi$.
To analyze the large radius limit it is convenient to rewrite the $C_l$ in a way that makes it explicit that $A_{\phi}$ is being sourced by a point mirror charge together with a mirror line charge extending from the mirror point charge to the origin:
\be
C_l = \frac{a^{l}}{d^{l+1}} \left [ g_2 + \frac{c_1}{-t_1 + l +1} + \frac{c_2}{-t_2+l+1} \right ].
\ee
Explicit formulas for point charge $g_2$ and the parameters $t_{1,2}$ and $c_{1,2}$ characterizing the line charge appear in the supplemental material of Ref.~\onlinecite{2009Sci...323.1184Q} for $\epsilon=\mu=1$. Most importantly, the point charge $g_2$ goes to the planar value $g$ in the $d \rightarrow a$ limit. We have confirmed with Mathematica that the same decomposition holds for arbitrary $\epsilon$ and $\mu$. The general formulas for $g_2$, $t_{1,2}$ and $c_{1,2}$ are too lengthy to reproduce here. Taking
$a$ large at fixed $d$, one sees that every term in the potential scales as $1/a$ and so the potential
seems to vanish
in the large $a$ limit. This conclusion is too fast as the infinite sum can alter the behavior. At $\theta=0$ (where $P_l(1)=1$ for all $l$) the sum can be performed analytically and we see that the $r=a$ contribution for the point charge
sums up to the expected $g_2/(d-a) = g_2/z_0$, that is it remains finite in the large $a$ limit. As the sum over Legendre polynomials by construction just represents a standard $1/r$ Coulomb potential, the point charge term automatically reproduces the planar contribution to $\theta_S$. At $\theta=0$ the contribution of the line charge can also be summed up. The resulting Hypergeometric function vanishes as $1/a$ in the large $a$ limit. As $P_l <1$ for $\theta \neq 0$ it is clear that at non-zero $\theta$ the line charge contribution has to vanish at least as fast as for $\theta=0$. So in the $a \rightarrow \infty$ limit, the line charge contribution can be neglected compared to the point-like one. This result has also been confirmed numerically in Ref.~\onlinecite{2009Sci...323.1184Q}.\\
 This shows that for loops of size $l \ll a$ we recover the planar result. However in the planar case we found that for $l \gg z_0$ the exchange is governed by the universal $\theta_S$. From this we conclude that for a compact geometry it is the intermediate size loops with
$z_0 \ll l \ll a$ which are governed by the universal topological phase of eq.~\eqref{thetas}. While we only explicitly demonstrated this result in the case of a sphere, we believe it to be true in general.

\section{Discussion}

We find zero
angular momentum for a charge outside a spatially compact TI in two examples. While obvious from the microscopic point of view
of electrons and protons, the result looks somewhat surprising starting from Maxwell's equations of a topological
insulator; in fact we haven not been able to give an analytic proof based on this effective theory that the total angular momentum always vanishes in a finite system even though the microscopic point of view strongly suggest that this is true.\\
By looking at the two particle system we find that the universal statistical phase governs the behavior of intermediate size loops.
In realistic experiments, our results indicate that to observe such an anyon in the
topological insulator set-up as suggested in Ref.~\onlinecite{2009Sci...323.1184Q}, one should use a large bulk sample with a small superconductor loop close to it, to get non-zero flux. The size of the superconductor loop has to be in the intermediate regime we identified, that is much smaller than the size of the sample but much large than the distance between loop and sample. Our results are consistent with the 2+1 dimensional nature of the anyons, as the universal statistical angle ceases to accurately describe the two particle systems once the particles are removed from the surface of the TI beyond a distance of order the sample size.

\begin{acknowledgments}
We would like to thank E.~Witten as well as X.-L.~Qi, R.~Li, J.~Zang and S.-C.~Zhang for very helpful suggestions.
The work of S.~Sun was supported in part by U.S. DOE grant No. DE-FG02-00ER41132, the work of A.~Karch by U.S. DOE grant No. DE-FG02-96ER40956.

\end{acknowledgments}

\begin{appendix}
\section{Derivation of the statistical angle in SI units}

Consider a planar interface between a topological
insulator (with non-trivial $\mu_2$ and $\epsilon_2$ as well as $\theta=\pi$,
that is $P_3=1/2$)
and a trivial insulator (with $\epsilon_1$ and $\mu_1$).
Consider a single static point charge $q$ inside the trivial
material a distance $z_0$ away from the surface
of the TI.

We want to find the mirror charges for general $\mu$ and $\epsilon$.
Let the TI occupy the $z<0$ region of space.
Writing Maxwell's equation as usual as
$$
\vec{\nabla} \cdot \vec{D} =  \rho_e,
\vec{\nabla} \times \vec{H} =
\frac{\partial \vec{D}}{\partial t} +  \vec{j}_e,
\vec{\nabla} \cdot \vec{B} =  \rho_m,
\vec{\nabla} \times  \vec{E} =
\frac{\partial \vec{B}}{\partial t} +  \vec{j}_m.
$$
The constitutive relations in SI unites are
$$\vec{D} = \epsilon \vec{E} -  \epsilon_0 \alpha \frac{\theta}{\pi} \, ( c_0 \vec{B}),
\quad \quad
c_0 \vec{H} = \frac{c_0 \vec{B}}{\mu} + \alpha \frac{\theta}{ \pi} \frac{\vec{E}}{\mu_0}.
$$

We can introduce potentials
$\Phi_{e,m}$ with
$\vec{E} = - \vec{\nabla} \Phi_e$ and $\vec{B} = - \vec{\nabla} \Phi_m$.
Above the interface they are given by (note that we weight all electric mirror charges by $\epsilon_0$ for convenience; we also use $\epsilon_1$ for $q_e$ on both sides of the interface; these are just definitions of our mirror charges)
$$ \Phi_e^I = \frac{q_e}{4 \pi \epsilon_1 R_1} + \frac{q^{(2)}_e}{4 \pi \epsilon_0 R_2}, \quad
\Phi_m^I = \frac{q^{(2)}_m}{ 4 \pi R_2}
$$
and below by
$$
 \Phi_e^{II} = \frac{q_e}{4 \pi \epsilon_1 R_1} + \frac{q^{(1)}_e}{ 4 \pi \epsilon_0 R_1},
\quad
\Phi_m^{II} =  \frac{q^{(1)}_m}{4 \pi R_1}
$$
where $q_{e,m}^{(1,2)}$ are the mirror charges locate a distance $d$ above
(for $q_{e,m}^{(1)}$) or below (for $q_{e,m}^{(2)}$) the interface.
$R_1^2=x^2+y^2+(z_0-z)^2$
and $R_2^2 = x^2 + y^2 + (z_0+z)^2$.
Maxwell's equations
in the absence of surface currents or charges as usual demand continuity of
$D_{\perp}$, $B_{\perp}$, $H_{\|}$ and $E_{\|}$.
As, at $z=0$, $R_1=R_2$, $\partial_z R_1 = - \partial_z R_2$ this demands:
\begin{eqnarray*}
(q_e- \frac{\epsilon_1}{\epsilon_0} q_e^{(2)})
&=& (\frac{\epsilon_2}{\epsilon_1} q_e+\frac{\epsilon_2}{\epsilon_0} q_e^{(1)}) - \epsilon_0 \alpha \frac{\theta}{\pi} (c_0 q_m^{(1)}) \\
q_m^{(1)} &=& - q_m^{(2)} \\
\frac{q_m^{(2)}}{\mu_1}
&=& \frac{q_m^{(1)}}{\mu_2} + \alpha \frac{\theta}{\pi} \frac{q_e/\epsilon_1 + q_e^{(1)}/\epsilon_0}{\mu_0 c_0} \\
q_e^{(2)} &=& q_e^{(1)} .
\end{eqnarray*}
From this the mirror charges can easily be found:
$$q_m^{(2)} = - q_m^{(1)} = \frac{1}{c_0} \frac{2 \alpha \frac{\theta}{\pi} q}
{(\epsilon_1 + \epsilon_2)(\frac{\mu_0}{\mu_1} + \frac{\mu_0}{\mu_2}) +
\epsilon_0 \alpha^2 \frac{\theta^2}{\pi^2}}$$
and
$$
q_e^{(2)} = q_e^{(1)} = \frac{\epsilon_0}{\epsilon_1} \frac{(\epsilon_1-\epsilon_2)
(\frac{\mu_0}{\mu_1} + \frac{\mu_0}{\mu_2}) - \epsilon_0 \alpha^2 \frac{\theta^2}{\pi^2}}{
(\epsilon_1 + \epsilon_2)(\frac{\mu_0}{\mu_1} + \frac{\mu_0}{\mu_2}) +
\epsilon_0 \alpha^2 \frac{\theta^2}{\pi^2}
} \, q
$$
The system
consisting a the charge $q$ and a magnetic charge $q_m$ gives rise to an angular momentum
$$ \vec{L} = \frac{q q_m}{4 \pi} \hat{r}$$
where $\hat{r}$ is the unit vector pointing from the electric to the magnetic charge. Clearly $\vec{L}$ vanishes when the two charges are sitting on top of each other (as in this case $\vec{E}$ and $\vec{B}$ are parallel, so the momentum density and hence the angular momentum density vanish identically).

For the interface, we need to calculate the contributions to the angular momentum in the two regions independently. Inside the TI both electric and magnetic fields are pointing radially outward from the point at $z=+z_0$, so the angular momentum vanishes (again, $\vec{E}$ and $\vec{B}$ are parallel and so the Poynting vector vanishes identically). For the region above the interface, we get a non-zero contribution to the angular momentum due to the charge/monopole system formed by the original charge $q$ at $z=+z_0$ and the mirror magnetic charge $q_m^{(2)}$ at $z=-z_0$ (the electric mirror charge $q_e^{(2)}$ at $z=-z_0$ does not contribute, as the charge at $z=+z_0$ is purely electric). If we calculate the angular momentum with respect to the origin $x=y=z=0$, we see that the integrand
\beq \nonumber & \vec{r} \times (\vec{E} \times \vec{B}) \sim \frac{ \vec{r} \times [ (\vec{r} - z_0 \hat{e}_z) \times (\vec{r} + z_0 \hat{e}_z) ]}{| \vec{r} - z_0 \hat{e}_z|^2 |\vec{r} + z_0 \hat{e}_z|^2}\\
& =
z_0 \, \frac{ \vec{r} \times [ \vec{r}  \times \hat{e}_z ]}{| \vec{r} - z_0 \hat{e}_z|^2 |\vec{r} + z_0 \hat{e}_z|^2}\eeq
is symmetric under $\vec{r} \rightarrow - \vec{r}$. So we get equal contributions to the angular momentum from the lower and the upper half plane. As in our case we only get a contribution from the upper half plane, the angular momentum to the charge/mirror-charge system is exactly {\it half} of what it would be for a genuine charge/monopole pair, that is (setting $\theta=\pi$, $q=-e$ in the expressions above)
\beq \nonumber & L_z = - \frac{ q q_m^{(2)}}{8 \pi} =-\frac{1}{4 \pi c_0} \frac{\alpha e^2}
{(\epsilon_1 + \epsilon_2)(\frac{\mu_0}{\mu_1} + \frac{\mu_0}{\mu_2}) +
\epsilon_0 \alpha^2}\\
& =- \frac{\alpha^2 \hbar}
{(\frac{\epsilon_1}{\epsilon_0} + \frac{\epsilon_2}{\epsilon_0})(\frac{\mu_0}{\mu_1} + \frac{\mu_0}{\mu_2}) +
\alpha^2}
\eeq
where in the last step we used the standard SI definition for $\alpha$,
$$
\alpha=\frac{e^2}{4 \pi \epsilon_0 \hbar c_0}.
$$

Note that for a given $q$, say $q=-e$, but for general $\mu$ and $\epsilon$ this is certainly not going to obey the Dirac quantization condition. The statistical angle one would want to associate with a dyon with angular momentum $L_z$ is $\theta_S =2 \pi L_z/\hbar$, so that the statistical angle $\theta_S$ is 0 for integer spins (bosons) and $\pi$ for half-integer spins (fermions). For a charge $-e$ in the presence of a TI surface we therefore obtain
$$|\theta_S| = 2 \pi \frac{\alpha^2}
{(\frac{\epsilon_1}{\epsilon_0} + \frac{\epsilon_2}{\epsilon_0})(\frac{\mu_0}{\mu_1} + \frac{\mu_0}{\mu_2}) +
\alpha^2}.$$

\end{appendix}

\bibliography{sphericalttNotes}
\end{document}